# Cable Fault Monitoring and Indication: A Review

[1] Shweta Gajbhiye, [2] S. P. Karmore

[1] Department of Computer Science and Engineering, GHRCE, Nagpur, Maharashtra, India

[2] Department of Computer Science and Engineering, GHRCE, Nagpur, Maharashtra, India

**Abstract**

Underground cable power transmission and distribution system are susceptible to faults. Accurate fault location for transmission lines is of vital importance. A quick detection and analysis of faults is necessity of power retailers and distributors. This paper reviews various fault locating methods and highly computational methods proposed by research community that are currently in use. The paper also presents some guidelines for design of fault location and remote indication, for reducing power outages and reducing heavy loss of revenue.

## 1. Introduction

In electrical utilities, transmission lines form the backbone of power systems. With regard to reliability and maintenance costs of power delivery, accurate fault location for transmission lines is of vital importance in restoring power services and reducing outage time as much as possible. Accurately locating faults on high voltage transmission networks is very important for utilities to allow a quick maintenance action of the repair crew.

Underground power cables have been widely implemented due to reliability and environmental concerns. To improve the reliability of a distribution system, accurate identification of a faulted segment is required in order to reduce the interruption time during fault, i.e. to restore services by determining a faulted segment in a timely manner. In the conventional way of detecting fault, an exhaustive search in larger scale distance has been conducted. This is time consuming and inefficient, Not only that the manpower resources are not utilized, but also the restoration time may vary depending on the reliability of the outage information. As such deriving an efficient technique to locate a fault can improve system reliability. POWER utilities need an accurate and automatic fault location method for ULVDN. This is due to number of key factors namely: reliability of supply, quality of supply, reducing operating costs of repairs and charging staff works practices, and low tariff charges to maintain a competitive edge. Path selection problems results unnecessary circumvent which in turn costs more in the overhead line construction. The trend of transmission line construction from overhead to underground is increasing even though the underground system costs more for initial construction. However, the underground system requires faster detection and correction of accidental faults along the lines for more reliable service.

Fault location detection is finding exact fault position of cable when there were any unwanted accidents like short circuit, open circuit, insulation breakdown etc. Because of large damage and inference of power cable accident, power authorities want to have exact fault detection method to recover power lines as soon as possible. Various methods have been developed to reduce damage and inference. But most of fault detection methods have shortcomings. Some have low accuracy, some are difficult to apply because of surrounding environment, and some give unwanted damage to healthy neighboring cable and facilities [3-4].

Among these methods, pulse echoing method is regarded as most useful ones. This method use time difference between incident and reflected pulse to calculate fault location detection and it has relatively high accuracy because it use short period pulse. Although it has high accuracy, pulse echoing method has some drawbacks. When we apply this method to low impedance accident, the error will be increased. If cable is not open circuit and there is no impedance change, there are no reflected pulse waves, and it is difficult to find fault location. Also high voltage pulse generator is needed for pulse echoing method and it is one of its drawbacks. To make high voltage pulse generator one should pay another money, and inserting high voltage to cable can cause another damage to cable and facilities. It is usually applied to detect fault location after accident arises and this way need more time to repair.

Because it is very expensive system, on-line monitoring and fault location detection of cable require so big money. And its natured drawbacks mentioned above, power authorities needs new methods which can detect fault location exactly and which on-line monitoring is possible.

## 2. Types of Cable Faults

Cable faults can be categorized into three main types : Open conductor faults, shorted faults, and high impedance faults [1-2].

2.1 Open Conductor Fault

An open conductor fault is where the conductor of a cable is completely broken or interrupted at the location of the cable fault. It is possible to have a high resistance shunted faults(to ground) on one or both sides of the faulted conductor's location.

2.1 Shorted Fault

A shorted fault is characterized by a low resistance continuity path to ground (shunted fault). The resistance from the conductor to ground is lower than the surge impedance of the cable for a shorted low resistance fault.

2.3 High Impedance Fault

A high impedance fault contains a resistive path to ground (shunted fault) that is large in comparison to the cable's surge impedance. This fault type may also demonstrate non-linear resistive characteristics which allow the apparent resistance to vary with the level of applied voltage or current.

## 3. Types of Faults Detection

The faults occurring in the power lines and cables can be classified into four main categories- short circuit to another conductor in the cable, short circuit to earth, high resistance to earth and open circuit.

Not all approaches work best for each type of fault. Four methods that are mostly used in detecting fault location are described as follows.
- A frame
- Thumper
- Time Domain Reflectometer (TDR)
- Bridge methods

A persistent short circuit to earth fault can be most easily located using A-Frame method. For high resistance to earth faults. A-Frame method is not always sufficient. In this case, thumper method needs to be used to reduce fault resistance. Thumper method alone may be sufficient for fault location but when applied for a longer duration, it may damage the cable insulation. A-Frame is not useful for locating faults which do not have an earth connection. Time Domain Reflectometer (TDR) is suitable for determining the locations of most of the faults.

3.1 A-Frame Method

In A-Frame method, a pulsed direct current (DC) is injected into the faulty cable and earth terminal to locate the ground fault. The DC pulse will flow through the conductor and return via earth from the earth fault location back to the ground stake as shown in Figure 1. The flow of pulsed DC through the ground will produce a small DC voltage. A sensitive voltmeter is used to measure the magnitude and direction of the DC voltage in segments of the earth along the cable route. Analyzing the results of the measuring voltage along the route, the location of the fault in the cable can be pinpointed A-Frame is an accurate method but it is not the fastest one, since the operator has to walk along the length of the cable from the transmitter to the ground fault.

This method may face a problem if the return DC finds some easier path back to the earth stake of transmitter instead of returning through the ground. If the ground is sandy, paved which provides high resistance and consequently, less current flows through the ground. In that case, the voltmeter fails to measure the voltage and fault detection becomes complicated.

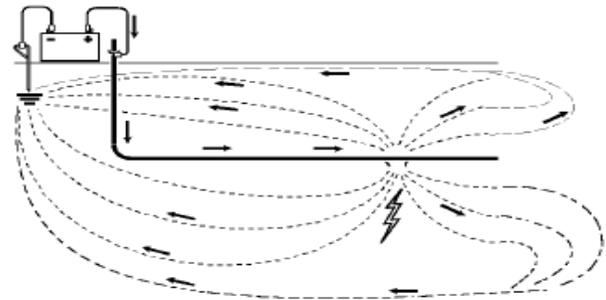

Figure 1 An A-Frame method of finding cable fault location

3.2 Thumper Method

Thumper is basically a high voltage surge generator which is used to apply a reasonable high voltage to the faulty core of an underground cable to generate a high current arc resulting in a loud noise to hear above the ground. This method requires very high current thump at voltages as high as 25 kV to make underground noise loud enough to be loud enough to be heard from the ground.

In thumper method of finding fault locations Like A-Frame, the thumper method requires an operator needs to walk along the path of the cable and listen for the sound from above the ground. Different ground conditions, nearby traffic and noises may make this sound hard to listen to make a clear distinction.

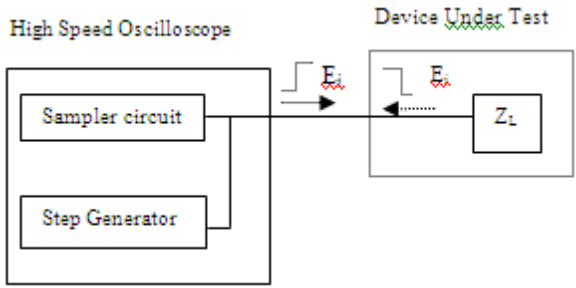

Figure 2 Functional block diagram of a TDR

### 3.3 Time Domain Reflectometry (TDR)

In the Time Domain reflectometry (TDR) method, a low energy signal is sent through the cable where the perfect cable with the uniform characteristic impedance returns the signal within a known time and with a known profile. This time and profile of the signal is altered once the cable has impedance variation due to any fault. The impedance variation causes a portion of the signal reflected back to source. The reflected signal fortifies the original signal when there is an increase in characteristic impedance at the fault location, while it opposes the original signal when there is a decrease in characteristic impedance. Graphical representation on the Time Domain Reflectometry (TDR) screen gives the user the distance to the fault in time units. The actual distance can be calculated by multiplying the time by signal velocity. The functional block diagram of a TDR is given in Figure 3.

Therefore, the low voltage TDR and the thumper methods can be integrated into a single system where a low voltage TDR pulse is taken of the cable under test and stored in a display memory. Then the thumper can be used to send a high voltage pulse for burning the faulty point. While the arc is burning at the faulty point, the TDR can be used to send the same low voltage pulse and new pulse will be superimposed upon the first trace. The arc is low impedance point that results in TDR pulse to reflect as it would with a short circuit. Figure 4 shows an example of the test with two traces of the signals one on the top of another.

In the figure, the dashes cursor represents the launching point and solid cursor shows the faulty point. From these two cursors, the machine can directly calculate the distance of the fault. The integrated thumper and TDR method reduces the major insulation damage of the cable but does not discard the risk. TDR method is useful for open circuit fault detection. Again if it has a low series resistance at the fault the problem will be similar as high resistance earth fault [8-9].

### 3.4 Bridge Method

Bridge methods used for locating faults in underground cables are based on modified Wheatstone circuit where direct current is used to measure the resistance in order to calculate distance of the fault in percentage of the total line length. Murray and Glaser bridges [1] use the similar principal for calculating the distance of the fault. Brief description of these bridges is given as follows.

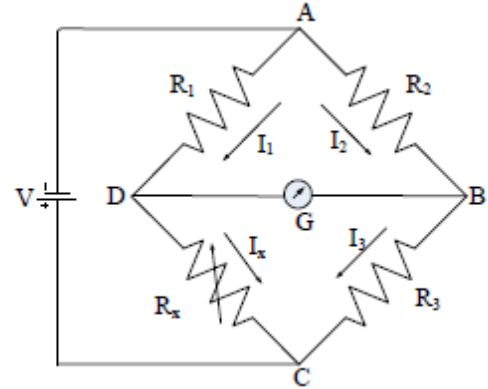

Figure 3 Wheatstone bridge circuit.

Figure 3 shows the Wheatstone bridge circuit where R1,R2,R3 are the known resistors and Rx is unknown resistor. When the galvanometer represented by the circle in the figure shows zero current flow, the unknown resistor Rx value can be found from the other known resistor value using following equation

$$R_x = \frac{R_2 \times R_3}{R_1} \qquad (1)$$

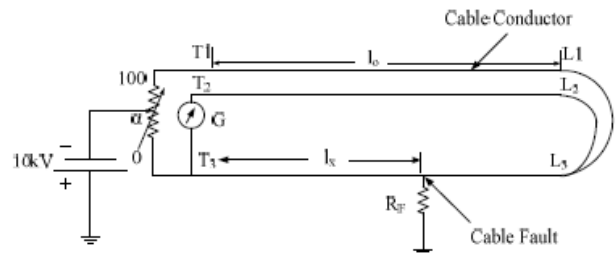

Figure 4 Murray bridge loop for cable fault location

Figure 4 shows a Murray bridge loop for cable fault location. Besides the faulty conductor, one healthy conductor is required as connected to terminal $T_1$ in the measuring circuit. External loop wires in the circuit, connecting the resistances at the front and the conductors at the cable end, should have close to zero resistance. In Murray Bridge loop, for the balance condition of the galvanometer. The equations for calculating the conductor resistances to the fault location are given below.

$$R_1 = \frac{R \times R_3}{R_3 + R_4} \quad (2)$$

$$R_1 = \frac{R \times R_4}{R_3 + R_4} \quad (3)$$

Where $R_1 + R_2 = R$

Fault distances can be calculated from the resistor values using the conductor resistance per unit length of cable.

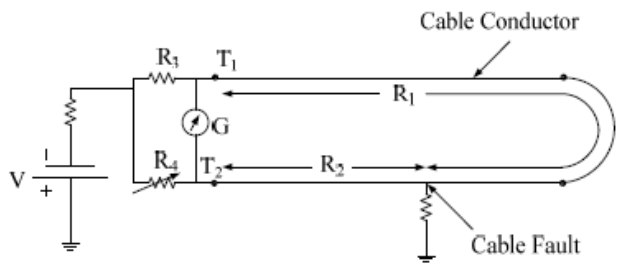

Figure 5 Glaser measuring circuit.

Glaser measuring circuit for cable fault locations is shown in Figure 5. In this measurement circuit, two healthy conductors ($L_1$ and $L_2$) of the cable with same diameter and material are required beside the faulty conductor ($L_3$) As shown in Figure 4. For the balanced condition of the circuit, the formula for finding cable fault distance is given below.

$$l_x = \frac{\alpha}{100} \times l_0 \quad (4)$$

Where $l_x$ is the distance of the fault from measuring end, $\alpha$ is the resistance up to the supply connection and $l_o$ is the length of the cable.

## 4. Types of Faults Detection

Here various fault locating methods are presented, that have been proposed by various researchers and yet to be tested on practical field.

H. Shateri, S. Jamali *Et Al.,* Proposed An impedance based fault location method for phase to phase and three phase faults[9] . This method utilized the measured impedance by distance relay and the super imposed current factor to discriminate the fault location. This method is sensitive to the measured impedance accuracy and super imposed current factor.

Abhishek Pandey, Nicolas H. Younan, Presented underground cable fault detection and identification vis fourier analysis[10]. The methods of impedance calculation via sending end voltage and differential voltage can be used for differentiating between the different types of cable defects from phase information. It needs study to be conducted to find the best way of visualizing the results, especially the magnitude response.

A.Ngaopitakkul, C. Pothisarn, M. Leelajindakrairerk[11], presented behaviour of simultaneous fault signals in distribution underground cable using DWT. The simulations were performed using ATP/EMTP, and the analysis behaviour of characteristics signals was performed using DWT. Various case studies have been carried out including the single fault and simultaneous fault.

Yuan Liao, Ning Kang [12] has presented fault location algorithms without utilizing line parameters. By utilizing unsynchronized voltage and current measurements from both ends of line without requiring line parameters based on the distributed parameter line model. The fault location estimatie is not sensitive to measurement errors while line parameter estimates are sensitive to measurement errors. Thus relatively precise measurements are required to obtain accurate line parameter estimates.

S. Navaneethan, J. J. Soraghan, W. H. Siew, F. McPherson, P. F. Gale [13] , presented an automatic fault location method using TDR. This method uses acquired data from an existing TDR instrument. It enables user of TDR equipment to locate ULVDN cable faults without user interpretation.

These fault locating methods describe in this sections are mostly computational. The authors of this paper found no evidence that they are in use in practical field to pinpoint the cable faults. However, these methods are mostly suitable for comparing the performances of distant/impedance based operating relays for their operational accuracy.

## 5. Conclusions

This paper explains the importance of locating faults in the distribution network and reviews some of the cable fault locating methods that are mostly used in practical field. There is a need to immediate indication about occurrence of a fault via remote communication; hence it needs to implement some techniques which will help power utilities in immediate indication of fault occurrence and accurate methods for locating faults. To facilitate the development, the preliminary investigation requirements and the essential segments to be verified are presented in this paper.